\documentclass[twocolumn]{svjour3}          
\smartqed  
\usepackage{graphicx}
\usepackage{mathptmx}      
\usepackage{braket}
\usepackage{cite}
\usepackage[utf8]{inputenc}
\usepackage{amssymb}
\usepackage{amsmath}
\usepackage{hyperref}
\usepackage{mathdots}
\usepackage{placeins}
\usepackage{bm}
\usepackage{subfigure}

\newcommand{\mean}[1]{\langle{#1}\rangle}

\begin{document}

\title{Overfitting in quantum machine learning and entangling dropout
}


\author{Masahiro Kobayashi$\mbox{}^1$ \and
	Kouhei Nakaji$\mbox{}^{2,3}$ \and 
	Naoki Yamamoto$\mbox{}^{1,2,\star}$
}


\institute{
       1, 
       Department of Applied Physics and Physico-Informatics, Keio University, Hiyoshi 3-14-1,
       Kohoku, Yokohama 223- 8522, Japan \\
       2, 
       Quantum computing center, Keio University, Hiyoshi 3-14-1,
       Kohoku, Yokohama 223- 8522, Japan \\
       3, 
       Present address: Research Center for Emerging Computing Technologies, National Institute of Advanced 
       Industrial Science and Technology (AIST), 1-1-1 Umezono, Tsukuba, Ibaraki 305-8568, Japan \\
       $\star$ Email: yamamoto@appi.keio.ac.jp
}


\maketitle


\begin{abstract}

The ultimate goal in machine learning is to construct a model function that has a generalization 
capability for unseen dataset, based on given training dataset. 
If the model function has too much expressibility power, then it may overfit to the training data and 
as a result lose the generalization capability. 
To avoid such overfitting issue, several techniques have been developed in the classical machine 
learning regime, and the dropout is one such effective method. 
This paper proposes a straightforward analogue of this technique in the quantum machine learning 
regime, the entangling dropout, meaning that some entangling gates in a given parametrized quantum 
circuit are randomly removed during the training process to reduce the expressibility of the circuit. 
Some simple case studies are given to show that this technique actually suppresses the overfitting. 

\end{abstract}

\keywords{Quantum machine learning \and parametrized quantum circuit \and overfitting 
\and dropout regularization}



\section{Introduction}

Quantum computer has potential to execute machine learning tasks more efficiently than 
conventional computers \cite{Biamonte2017,Mitarai2018,havlivcek2019}. 
Both in classical and quantum machine learning problems, it is important to use an approximator 
(a model function) that has a high expressibility power to achieve a good computational accuracy. 
For this purpose, recently the data re-uploading method was proposed 
\cite{Vidal2020,Salinas2020,Schuld2021}, 
which encodes the input data into the quantum circuit multiple times in the depth direction of the 
circuit. 
On the other hand, if the expressibility is too high, then the approximator may overfit to the 
training dataset and lose the generalization capability for unseen dataset. 
That is, we need to carefully design an approximator with appropriate expressibility power, 
such that it can achieve as small generalization (out-of-sample) error as possible. 
In fact, we recently find some theoretical analysis on the generalization error in quantum machine 
learning regime \cite{Caro2021,Gyurik2021,banchi2021,chen2021,Du2021,Caro2021b}, 
although practical quantum-oriented methods for suppressing the overfitting issue have not 
been sufficiently discussed.

In this paper, we propose a practical method for suppressing the overfitting issue in quantum 
machine learning problems, the {\it entangling dropout}. 
This is a straightforward analogue of the classical dropout \cite{Srivastava2014}, which randomly 
removes some connections between nodes of a neural network during the training process; 
although the mechanism for suppressing the overfitting has not been fully revealed, the effectiveness 
of classical dropout is very well recognized, and it has been widely used. 
The proposed entangling dropout is a similar technique, that randomly removes some entangling 
gates contained in a given parametrized quantum circuit during the training process. 
This clearly decreases the expressibility, but its actual effectiveness is not obvious; 
in this paper some examples are given to show that the entangling dropout certainly works to 
suppress the overfitting and improve the out-of-sample errors. 
We then discuss how to choose the dropout ratio, which determines the number of entangling 
gates removed in each iteration. 
Also, a comparison to the $L_1$ or $L_2$ regularization technique is provided, showing 
a merit of using the entangling dropout method.

Note that recently another type of dropout for quantum machine learning was proposed 
\cite{Verdon2018,Schuld2020}; this scheme randomly removes a qubit, by directly 
measuring it and discarding the result (i.e., taking the partial trace) followed by adding another 
qubit to the same place. 
Hence this scheme may be termed the qubit dropout. 
Clearly this operation completely blocks the information flow on the quantum circuit, reflecting 
a more straightforward analogue to the classical dropout. 
On the other hand, the entangling dropout is easier to implement, compared to the qubit 
dropout that needs an intermediate measurement on a circuit. 
Comparing the effect of these two type of regularization methods is interesting, but in this paper 
we only investigate the proposed entangling dropout.


\section{Overfitting in quantum machine learning}

\subsection{The quantum circuit model}

We use the following $n$-qubits quantum circuit to construct an approximator; 
for an input vector $x$, the scalar-valued output is given by 
\begin{align}
\label{output function}
    & f_L(x ; \theta) = \braket{0^n|U_L^{\dagger}(x ; \theta)AU_L(x ; \theta)|0^n}, \\ \nonumber
    & U_L(x ; \theta) = W(\theta^{(L)})S(x)W(\theta^{(L-1)}) S(x) \cdots W(\theta^{(1)}) S(x), 
\end{align}
where $A$ is an observable and $|0^n\rangle=|0\rangle^{\otimes n}$ with $|0\rangle$ 
the computational 0 state. 
$W(\cdot)$ is the unitary matrix representing a parametrized quantum circuit, where 
$\theta=(\theta^{(1)}, \ldots, \theta^{(L)})$ is the set of tunable parameters with 
$\theta^{(\ell)}$ the vector of parameters contained in the $\ell$th $W(\cdot)$. 
$S(x)$ is a circuit for encoding $x$, which is composed of the unitaries of the form $e^{-ig(x)H}$ 
in the 1-dimensional case $x\in{\mathbb R}$ with $g(x) \in{\mathbb R}$ a function we can arbitrarily 
choose. 
It is then shown in Ref.~\cite{Schuld2021} that $f_L(x ; \theta)$ can be expressed as the following 
Fourier form 
\begin{equation}
\label{fourier expansion}
    f_L(x ; \theta) = \sum_{\omega \in \Omega}c_{\omega}(\theta)e^{i\omega g(x)}, 
\end{equation}
where $\Omega$ denotes the discrete set of frequencies. 
The number of elements of $\Omega$ is an increasing function of $L$, implying the importance 
of multiple use of the encoder $S(x)$ for the purpose of enhancing the expressibility power of 
$f_L(x ; \theta)$.

\subsection{Example of overfitting}

\begin{figure}[tb]
     \includegraphics[width=\columnwidth]{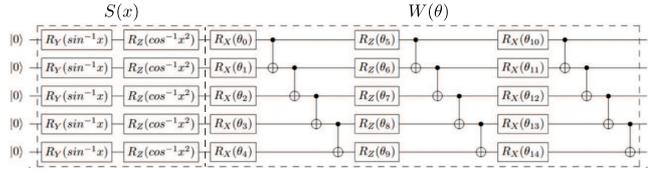}
     \caption{Diagram of the encoding circuit $S(x)$ and the tuning circuit $W(\theta)$ in the case $n=5$.}
     \label{fig-circuit diagram}
\end{figure}

Let us consider a simple 1-dimensional regression problem; given a training dataset 
$\{(x_k, y_k)\}_{k=1,\ldots,D}$ with $x_k\in{\mathbb R}$, the task is to determine $\theta$ 
so that the cost (in-sample error) 
\begin{equation}
\label{cost regression}
    J_{\rm r}(\theta) = \frac{1}{D} \sum_{k=1}^D \big( y_k -  f_L(x_k ; \theta) \big)^2
\end{equation}
is minimized. 
Here we take a $n=5$ qubit model, with the encoding circuit $S(x)$ and the tuning circuit 
$W(\cdot)$ illustrated in Fig.~\ref{fig-circuit diagram}; 
note that $S(x)$ is composed of $R_y(\sin^{-1}x)$ etc., representing $g(x)=\sin^{-1}x$ 
in Eq.~\eqref{fourier expansion}. 
The output is defined through $A=Z_1=Z\otimes I^{\otimes 4}$ with $Z$ the Pauli-$Z$ matrix 
and $I$ the $2\times 2$ identity matrix.

\begin{figure}[tb]
     \includegraphics[width=\columnwidth]{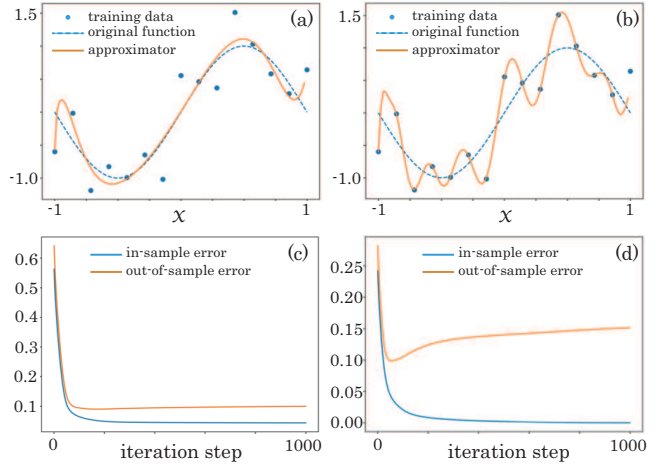}
     \caption{
     (a, b) Training dataset (blue points) and the generated approximator $f_1(x ; \theta)$ for 
     (a) and $f_{10}(x ; \theta)$ for (b). 
     (c, d) In-sample (training) error and out-of-sample (test) error achieved by $f_1(x ; \theta)$ for 
     (c) and $f_{10}(x ; \theta)$ for (d). }
     \label{fig-overfit example}
\end{figure}

Recall that the final goal of machine learning is to construct an approximator that has a good 
generalization capability, i.e., the function having a small out-of-sample error: 
\[
    J_{\rm r}'(\theta) = \frac{1}{D'} \sum_{k=1}^{D'} \big( y'_k - f_L(x'_k ; \theta) \big)^2, 
\]
where $\{(x'_k, y'_k)\}_{k=1,\ldots,D'}$ is the test dataset (or unseen dataset); 
note that $J_{\rm r}'(\theta)$ is not used to train $\theta$. 
If the approximator $f_L(x ; \theta)$ has too much expressibility, then $J_{\rm r}'(\theta)$ can 
take a large value even when $J_{\rm r}(\theta)$ is well suppressed, i.e., overfitting to the training 
dataset. 
This issue can actually be seen in the case where the training dataset is generated from 
$y=\sin(\pi x)$ with Gaussian noise; 
the dataset and $y=\sin(\pi x)$ are illustrated by the blue points and the blue dotted-line, respectively, 
in Fig.~\ref{fig-overfit example} (a), (b). 
We apply two models $f_1(x ; \theta)$ and $f_{10}(x ; \theta)$; as $n=5$, the number of parameters 
of these models are 15 and 150, respectively. 
Clearly, $f_{10}(x ; \theta)$ contains more frequency components (i.e., the size of $\Omega$ is bigger) 
and accordingly has bigger expressibility than $f_1(x ; \theta)$. 
After 1000 times update of parameters (we used Adam), $f_{10}(x ; \theta)$ overfits to the training 
dataset while $f_1(x ; \theta)$ looks a modest approximator, as shown in 
Fig.~\ref{fig-overfit example} (a), (b).

The overfitting issue can also be seen in the training process; 
Fig.~\ref{fig-overfit example} (c, d) show the in-sample error $J_{\rm r}(\theta_j)$ and the 
out-of-sample error $J_{\rm r}'(\theta_j)$ at the $j$th iteration step of parameters update, 
in the case (c) $L=1$ and the case (d) $L=10$. 
Note that $J_{\rm r}'(\theta_j)$ is computed with the learned parameter $\theta_j$ obtained 
using $J_{\rm r}(\theta)$. 
Clearly, as $J_{\rm r}(\theta_j)$ decreases, or equivalently the model becomes a better 
approximator to the training dataset, $J_{\rm r}'(\theta_j)$ for the case $L=10$ increases 
after at about $j=200$ iteration steps, while that of $L=1$ becomes almost constant. 
This is a well known phenomenon of overfitting that the out-of-sample error increases after 
some point of the parameter update, implying the necessity of early-stopping of the training 
which is also often used as a regularization technique for avoiding overfitting.

Lastly note that there have been some discussion stating that overfitting may not occur 
for the case $L=1$ \cite{Mitarai2018,chen2021}, which is actually consistent to the above 
numerical simulation result.


\section{Entangling dropout}

\subsection{Method}

\begin{figure}[tb]
     \includegraphics[width=\columnwidth]{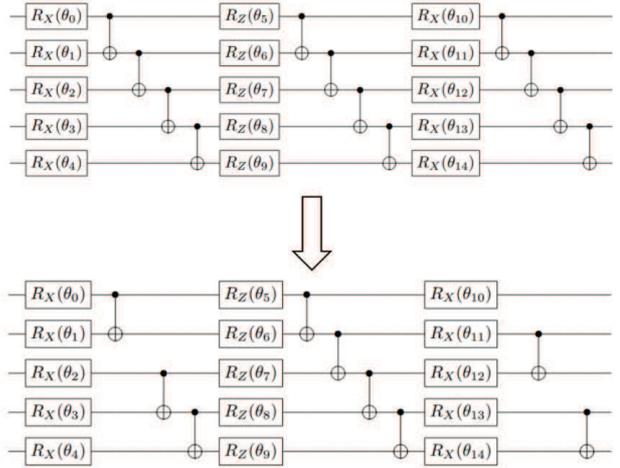}
     \caption{Example of removing CNOT gates with gate dropout rate 0.3, for the  chosen layer $W(\cdot)$.}
     \label{fig-dropout}
\end{figure}

The expressibility of a parameterized quantum circuit heavily depends on the number 
of entangling gates over qubits. 
Hence, similar to the dropout method for neural networks, we propose a method to reduce 
the expressibility by randomly removing some entangling gates contained in the circuit during 
the training process -- entangling dropout. 
A more precise description of the method is as follows: 
\begin{enumerate}
\item Remove some entangling gates randomly at the $j$th parameters update. 
\item Calculate the gradient of the cost function for the circuit with dropout executed, to 
determine the $(j+1)$th parameters using the gradient descent method. 
\item Repeat the above procedure until a desired performance is reached. 
\end{enumerate}
Note that the entangling dropout is executed independently at each iteration step, as in the 
classical case; hence, the average number of entangling gates is constant through the whole 
learning process. 
The random removal is determined by the following rule: 
\begin{enumerate}
\item Randomly choose layers in which some entangling gates are to be removed. 
The probability of choosing a layer to which the dropout is applied is called the {\it layer 
dropout ratio}. 
\item Then randomly remove entangling gates in the chosen layers. 
The probability of removing an entangling gate is called the {\it gate dropout ratio}. 
\end{enumerate}
These two hyper parameters should be carefully designed. 
For instance, let us consider the circuit with $n=5$ and $L=10$, and the case where the layer 
dropout rate and the gate dropout ratio are 0.2 and 0.3, respectively. 
In this case, on average, $10 \times 0.2 = 2$ layers are first chosen randomly and then, on average, 
$12 \times 0.3 = 3.6$ entangling gates in each chosen layer are randomly removed; 
Fig.~\ref{fig-dropout} shows the case where 3 CNOT gates are removed.

Remark:  In various simulations, we observed that the strategy defined with a single dropout rate for 
removing entangling gates in the whole circuit did not work so well; 
more precisely, we observed that, rather than removing entangling gates uniformly in the circuit, but 
some imbalanced assignment (e.g., some layers are completely free from the dropout operation) 
should be introduced to achieve a good performance. 
A related discussion will be given in Section~3.3.

\subsection{Suppression of overfitting}

\begin{figure}[tb]
\includegraphics[width=\columnwidth]{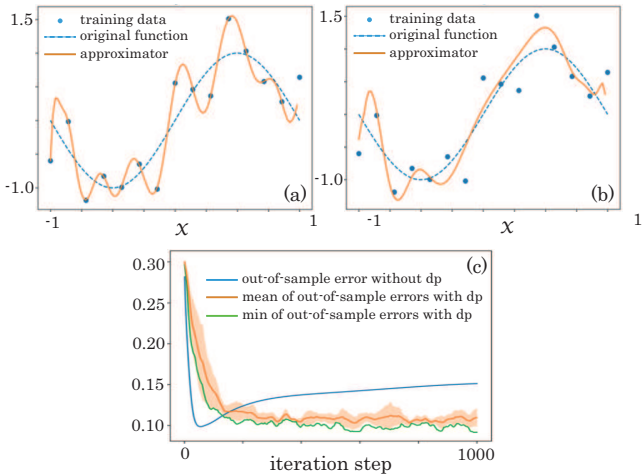}
     \caption{
     (a, b) Training dataset (blue points) generated from $y=\sin(\pi x)$ with Gaussian noise (blue 
     dotted line), and the constructed approximator $f_{10}(x ; \theta)$ without dropout (a) 
     and with dropout (b). 
     (c) Out-of-sample (test) errors achieved by $f_{10}(x ; \theta)$ without dropout (blue solid line) 
     and with dropout (mean: red solid line, min: green solid line). 
     The orange shade represents the standard deviation of 5 learning curves, fluctuating with 
     respect to the randomness in dropout. }
\label{first dp example}
\end{figure}

Here we demonstrate the performance of entangling dropout method. 
We continue to study the same regression problem discussed in Section 2.2, where both 
the training and test datasets are generated from $y=\sin(\pi x)$ with Gaussian noise; 
also the same circuit of $n=5$ qubits and $L=10$ layers is taken. 
The approximator $f_{10}(x ; \theta)$ constructed through the learning process is illustrated in 
Fig.~\ref{first dp example} (a), which is the same as  Fig.~\ref{fig-overfit example} (b); 
as seen before, because this approximator contains too many frequency components and 
accordingly have too much expressibility, overfitting occurs. 
Also, the orange line in Fig.~\ref{fig-overfit example} (d) is the same as the blue line in 
Fig.~\ref{first dp example} (c), showing that the out-of-sample error increases as the learning 
proceeds. 
Now we apply the entangling dropout with layer dropout rate 0.2 and gate dropout rate 0.2, 
to this circuit. 
Because the dropout is a stochastic operation, the constructed approximator differs in every trial; 
one example is illustrated in Fig.~\ref{first dp example} (b), showing that the approximator 
with dropout does not overfit to the training dataset, compared to that without dropout. 
As a result, the out-of-sample error does not increase, as illustrated in Fig.~\ref{first dp example} (c) 
showing the mean, standard deviation, and the minimum envelope of 5 sample paths of out-of-sample 
errors.

\begin{figure}[htbp]
     \includegraphics[width=\columnwidth]{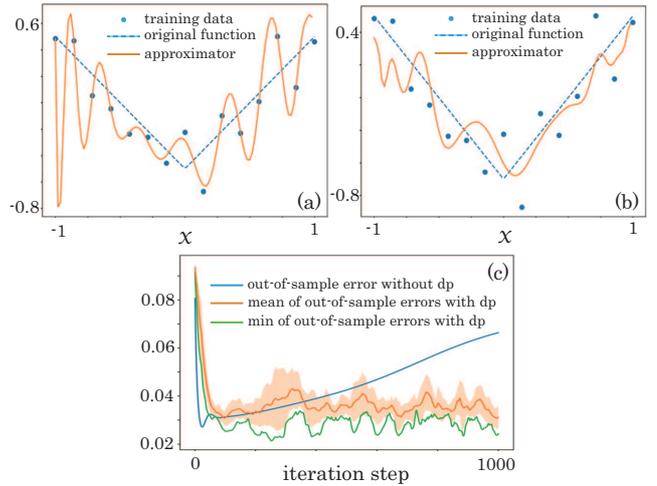}
     \caption{
     (a, b) Training dataset (blue points) generated from $y=|x|-1/2$ with Gaussian noise (blue 
     dotted line), and the constructed approximator $f_{10}(x ; \theta)$ without dropout (a) 
     and with dropout (b). 
     (c) Out-of-sample (test) errors achieved by $f_{10}(x ; \theta)$ without dropout (blue solid line) 
     and with dropout (mean: red solid line, min: green solid line). 
     The orange shade represents the standard deviation of 5 learning curves, fluctuating with 
     respect to the randomness in dropout. }
     \label{second dp example}
\end{figure}

To support the above observation emphasizing the usefulness of entangling dropout, we show 
another example of regression problem; the function generating the training and test dataset 
is $y=|x|-1/2$. 
The circuit is exactly the same as the previous case. 
The layer and gate dropout rate are 0.2 and 0.5. 
The result is summarized in Fig.~\ref{second dp example}, which may lead to the same conclusion 
reached above; that is, the dropout is effectively used for suppressing the overfitting and 
consequently the out-of-sample error does not increase.

\begin{figure}[htbp]
     \includegraphics[width=\columnwidth]{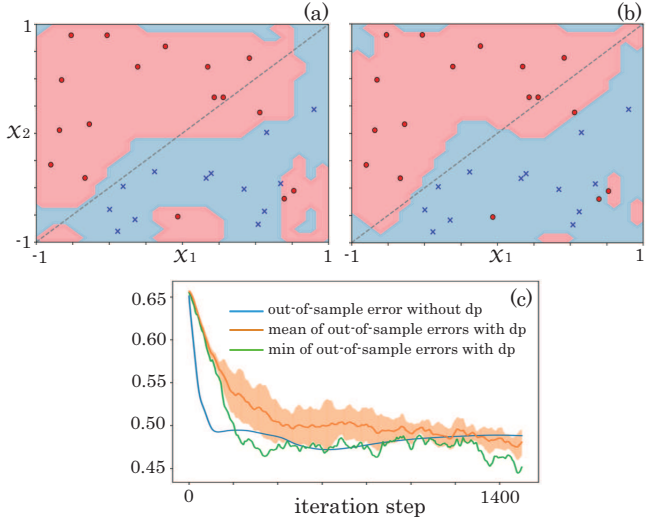}
     \caption{
     (a, b) Two-class training dataset is illustrated by the blue and red points. 
     The constructed classifier is illustrated as the boundary dividing the blue and red regions, 
     where (a) and (b) represent the case without dropout and with dropout, respectively. 
     (c) Out-of-sample (test) error achieved by the classifier without dropout (blue solid line) 
     and with dropout (mean: red solid line, min: green solid line). 
     The orange shade represents the standard deviation of 5 learning curves, fluctuating with 
     respect to the randomness in dropout. }
     \label{third dp example}
\end{figure}

It is also worth studying the effect of entangling dropout for a classification problem. 
Here we study the following 2-dimensional 2-label classification problem; given a training dataset 
$\{(x_k, y_k)\}_{k=1,\ldots,D}$ where the vector $y_k=(1, 0)$ or $(0, 1)$ represents the class 
assigned to the input $x_k\in{\mathbb R}^2$, the task is to construct a probabilistic classifier 
that assigns the label based on the probability vector 
\[
     F(x) = \frac{1}{e^{\mean{Z_1}}+e^{\mean{Z_2}}}
                  \left[\begin{array}{r}
                     e^{\mean{Z_1}}  \\
                     e^{\mean{Z_2}}  \\
                  \end{array}\right] 
            =  \left[\begin{array}{r}
                    1/(1+ e^{-f_L(x;\theta)}) \\
                    1/(1+ e^{f_L(x;\theta)}) \\
                \end{array}\right]. 
\]
Here $f_L(x ; \theta)$ is given in Eq.~\eqref{output function} with 
$A=(Z\otimes I - I\otimes Z)\otimes I^{\otimes (n-2)}$. 
The parameters $\theta$ are determined so that the cross entropy cost (in-sample error) 
\begin{equation*}
    J_{\rm c}(\theta) = \sum_{k=1}^D y_k^\top \log F(x_k)
\end{equation*}
is minimized ($\log$ is taken elementwise). 
We again consider the case $n=5$ and the same circuits $S(x)$ and $W(\theta)$ illustrated in 
Fig.~\ref{fig-circuit diagram}, with $L=10$. 
For the input $x_k=(x_{k,1}, x_{k,2})$, we encode $x_{k,1}$ into the the 1st, 3rd, 5th qubit while 
$x_{k,2}$ into the 2nd and 4th qubit. 
Then we compare the classifiers with and without entangling dropout; 
the layer and gate dropout rate are 0.2 and 0.3. 
The two-class training dataset is illustrated by the blue and red points in 
Fig.~\ref{third dp example} (a), (b). 
The constructed classifier is depicted as the boundary dividing the blue and red regions in 
Fig.~\ref{third dp example}, where (a) and (b) represent the case without and with dropout, 
respectively. 
Though not so visible, the classifier without dropout (a) looks like separating two regions with 
more complicated boundaries, compared to that with dropout (b). 
Figure~\ref{third dp example} (c) shows the out-of-sample errors over the training process of 
the classifier with and without dropout; recall that the dropout is a stochastic operation, and 
here the mean, standard deviation, and the minimum envelope of 5 sample paths of out-of-sample 
errors with dropout are shown. 
As in the regression case, the out-of-sample error without dropout increases (but slowly) after 
about 700 iteration steps, reflecting the overfitting, while the out-of-sample error with dropout 
does not show such a trend.

Remark:  It is interesting to see the effect of dropout for entangling gates other than CNOT gate. 
Our view is that using another fixed entangling gates such as $iSWAP$ gate or combinations of 
those gates will not greatly change the result of dropout. 
On the other hand, we expect that dropout on parametrized entangling gate will improve the 
resulting generalization performance; changing the parameters of an entangling gate in a quantum 
circuit can be interpreted as changing the activation rate of a neuron in a classical neural network, 
and, as for the latter, Ref.~\cite{Frey2013} demonstrated that a modified dropout method depending 
on such activity can improve the performance. 
Therefore, some modification on the entangling dropout method, which may introduce such adaptive 
rule depending on the status of parametrized entangling gates, will also improve the performance.

\subsection{Performance dependence on the dropout rate}

In this section, we study the effect of changing the layer and gate dropout rate, for the same 
regression problem in Section 2.2 and 3.2; that is, both the training and test datasets are 
generated from $y=\sin(\pi x)$ with Gaussian noise, and the circuit of $n=5$ qubits and $L=10$ 
layers is employed.

\begin{figure}[htbp]
     \includegraphics[width=\columnwidth]{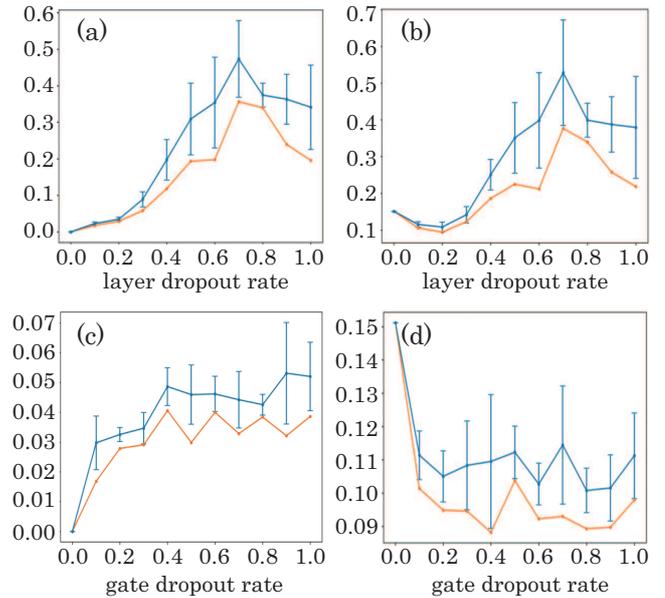}
     \caption{
     The in-sample error (a) and the out-of-sample error (b) as a function of the layer dropout rate, 
     where the gate dropout rate is fixed to 0.3. 
     Also the in-sample error (c) and the out-of-sample error (d) as a function of the gate dropout rate, 
     where the layer dropout rate is fixed to 0.2. 
     In the four figures, the blue solid line shows the mean of errors over 5 sample trajectories, 
     and the orange line is the minimum envelope of these trajectories. }
     \label{fig-dropout rate}
\end{figure}

Figure~\ref{fig-dropout rate} (a) and (b) show the in-sample error and the out-of-sample error, 
respectively, where the layer dropout rate is changed from 0.0 to 1.0 while the gate dropout 
rate is fixed to 0.3; the blue solid line shows the mean of errors over 5 sample trajectories, and 
the orange line is the minimum envelope of these trajectories. 
Because increase of dropout rate means decrease of expressibility of the approximator, the 
increasing trend of the in-sample error in the figure (a) makes sense. 
Notably, the out-of-sample takes the minimum when the layer dropout rate is 0.2; 
that is, if the circuit expressibility is too strong (i.e., the layer dropout rate is near 0), the approximator 
overfits to the training data and consequently the out-of-sample error becomes worse; 
also if the circuit expressibility is weak (i.e., the layer dropout rate is bigger than 0.3), it underfits 
to the training data and the same performance degradation is observed.

To see the possibility of further decreasing the out-of-sample error, we turn to examine the case 
where the gate dropout rate changes from 0.0 to 1.0 while the layer dropout rate is fixed to 0.2. 
Figure~\ref{fig-dropout rate} (c) and (d) show the in-sample error and the out-of-sample error, 
respectively (the meaning of blue and orange lines are the same as the case of (a, b)). 
The result is that, as shown in the figure (d), the gate dropout rate of 0.8 is the best choice 
in this case.

Hence, naively, the combination of small layer dropout rate and large gate dropout rate may lead 
to a good out-of-sample performance. 
This simple guide is supported by another case-study comparing two pairs of dropout rate shown 
in Table~\ref{table-dropout compare}. 
The point of this choice is that the average number of removed CNOT gates is almost the same 
in these two cases. 
Actually, the combination of small layer dropout rate (0.2) and large gate dropout rate (1.0) yields 
a better performance in the out-of-sample error, compared to the opposite combination of large 
layer dropout rate (0.6) and small gate dropout rate (0.3). 
Hence lessons learned here is that the entangling dropout should be executed with care in choosing 
the place in addition to the total number of entangling gate to be removed.

\begin{table}[htbp]
\caption{Comparison of two patterns of dropout (dp) rate}
\label{table-dropout compare}
        \centering
        \begin{tabular}{cccccc} 
          layer      &  gate     & average $\sharp$ of          & accuracy for & accuracy for   \\ 
          dp rate  &   dp rate  & removed CNOT   & training data & test data        \\ \hline
          0.2   & 1.0   & 24   & 0.052     & 0.11                  \\
          0.6   & 0.3   & 21.6   & 0.35       & 0.40               \\
      \end{tabular}
\end{table}


\subsection{Comparison to $L_1$ and $L_2$ regularization}
\label{change_dropout_rate}

Lastly we conduct a comparison to another regularization method, i.e., $L_1$ and $L_2$ regularization, 
which is also often used in conventional machine learning to suppress the overfitting issue. 
This technique is realized by simply modifying the cost function as follows: 
\begin{equation*}
      J(\theta) ~\rightarrow~ J(\theta) + \lambda \sum_{i}|\theta_{i}|^{p}, 
\end{equation*}
where the second term represents the penalty on freely choosing the parameters. 
The hyper parameter $\lambda$ is chosen to determine the weight of this penalty. 
$L_1$ and $L_2$ regularization correspond to $p=1$ and $p=2$, respectively. 
It is well known that, with this regularization, some parameters tend to have small value or 
become exactly zero, which thus reduce the number of effective parameters or equivalently 
the power of expressibility.

Here we again consider the same regression problem in Section 2.2 and 3.2, where the dataset 
is generated from $y=\sin(\pi x)$ with Gaussian noise and the circuit of $n=5$ and $L=10$ is 
chosen as the approximator. 
The result is summarized in Fig.~\ref{first dp example}; in particular the approximator without 
and with dropout reaches the out-of-sample error of about 0.1511 and 0.1092, respectively, 
after 1000 parameters update.

The comparable performance can actually be achieved via the $L_1$ and $L_2$ methods, 
by carefully choosing the wighting parameter $\lambda$. 
The result is listed in Table~\ref{table-regularization comparison}; actually they reach the 
value of out-of-sample error 0.1339 and 0.1483, which are smaller than 0.1511, yet are not 
smaller than 0.1092. 
Importantly, in this case $L_1$ and $L_2$ need 10000 steps of iteration, which is much 
bigger than the necessary number when using the entangling dropout.

\begin{table}[htbp]
\caption{Comparison of the entangling dropout and the $L_1$ or $L_2$ regularization method}
        \label{table-regularization comparison}
        \centering
        \begin{tabular}{cccccc}
          pattern    & accuracy for test data & number of iterations \\ \hline
          without dropout & 0.1511      & 1000 \\
          with dropout & 0.1092      & 1000 \\
          $L_1$          & 0.1339      & 10000 \\
          $L_2$          & 0.1483      & 10000 \\ 
        \end{tabular}
\end{table}

\begin{figure}[htbp]
\centering
\includegraphics[scale=0.31]{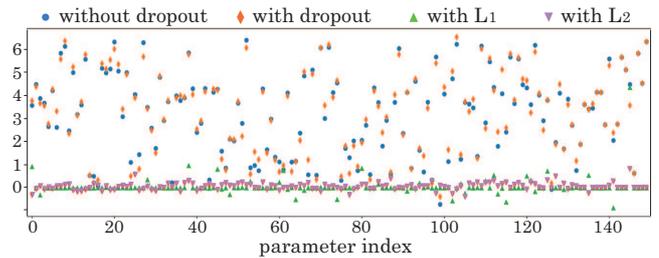}
\caption{The optimized parameters for the case of $L_1$, $L_2$, and with/without entangling dropout. 
The vertical axis represents the value of parameters.}
\label{fig-parameter info}
\end{figure}

It should be also worth studying the values of optimized parameters. 
Figure~\ref{fig-parameter info} shows those values, where the horizontal axis corresponds 
to the index of parameters. 
Interestingly, the optimized parameters obtained via the dropout method take the values 
(orange points) near the values without regularization (blue points); 
that is, the dropout method makes a slight change of parameters yet to have a large impact 
on the resultant out-of-sample error. 
This makes sense, because the dropout reduces the expressibility by directly limiting the flow 
of quantum information in the circuit, without largely changing the parameters of circuit. 
On the other hand, the $L_1$ and $L_2$ regularization keep the same reachable set in the 
original Hilbert space, and thus it carefully has to choose the parameters, which may explain 
the reason why 10000 iterations is necessary to learn the parameters.

\section{Conclusion}

In this paper, we proposed the entangling dropout method, for suppressing the overfitting 
issue in quantum machine learning problems. 
Some numerical demonstrations show that this method effectively achieves the goal, by 
directly reducing the power of entanglement and thereby limiting the reachable set in 
the feature Hilbert space. 
In general, designing an effective ansatz circuit in quantum variational algorithms is not 
an easy task; in this sense the entangling dropout can be interpreted as a method for randomly 
searching a proper ansatz suitable for quantum machine learning problems. 
We also note that the idea of entanglement dropout itself can be applied to various types of 
quantum machine learning problems, e.g., regression and classification problems where the 
dataset is given by a set of quantum states \cite{Biamonte2017,Osborne2020,Coles2022}.

The demonstrations shown in this paper are based on only a few small-size examples, and 
thus we cannot draw a general perspective on the entangling dropout method. 
Note that, in our view, it is not so important to perform many such small-size examples. 
Rather, the effectiveness of the entangling dropout method should be judged in future 
large-size quantum machine learning settings where hopefully some sort of quantum advantages 
would be established. 
Yet we believe it is worth sharing the idea of entangling dropout in the community, in advance 
to such demonstration.

\mbox{}
\\
{\bf Acknowledgements:} 
This work was supported by MEXT Quantum Leap Flagship Program Grants No. JPMXS0118067285 
and No. JPMXS0120319794, and also Grant-in-Aid for JSPS Research Fellow Grant No. 22J01501.

\mbox{}
\\
{\bf Conflict of interest:} 
On behalf of all authors, the corresponding author states that there is no conflict of interest.

\mbox{}
\\
{\bf Data availability statement:} 
All data generated or analysed during this study are included in this published article.

\end{document}